\documentclass{kluwer}    

\newdisplay{guess}{Conjecture}

\begin{document}                                                                                   
\begin{article}
\begin{opening}         
\title{3D Simulations of Galactic Winds in Dwarf Galaxies}
\author{Andrea \surname{Marcolini}}
\author{Fabrizio \surname{Brighenti}}
\runningauthor{Marcolini et al.}
\runningtitle{A Sample Document}
\institute{Department of Astronomy, University of Bologna}
\author{Annibale \surname{D'Ercole}}
\institute{Bologna Astronomical Observatory}

\end{opening}           

Galactic winds play a key role in the formation and evolution of galaxies
and the intergalactic medium (IGM). 
The fate of the heavy elements ejected by the SNII 
produced in a starbursting dwarf galaxy is a particular crucial issue: 
are the metals
blown out in the IGM and lost forever from the galaxy? Or are they 
confined by relatively large pressure gas existing at large radii
(both ISM or ICM)?
In the latter case  a significant fraction of the metals
might fall back into the galaxy, enriching the ISM.
In the simulation presented here we include a new ingredient 
affecting the evolution of galactic winds:
the ram pressure applied by the IGM on the ISM as the galaxy roams
through it. We consider a typical dwarf galaxy moving inside a typical
poor galaxy group. The model galaxy is built along the lines of 
D'Ercole \& Brighenti (1999, MNRAS, 309, 941), 
with a stellar disk with mass 
$M_*\sim 6 \times 10^8$ M$_\odot$ and a spherical dark halo 
$M_{\rm h} \sim 7\times 10^9 $ M$_\odot$. The ISM, in rotational 
equilibrium in the resulting potential, has mass 
$M_{\rm g} \sim 6\times 10^8 $ M$_\odot$ in the ``galactic region'',
defined as $|z|<1$ kpc, $R<2$ kpc, in cylindrical coordinates.
The galaxy moves with a velocity $v=200$ km s$^{-1}$ through an ICM 
with $n_{\rm e} = 10^{-4}$ cm$^{-3}$ and $T=10^6$ K.
After 1 Gyr, a central starburst generates a power 
$L=3.8\times 10^{40}$ erg s$^{-1}$, for 30 Myr. 
We follow the evolution of the
resulting galactic wind for 500 Myr.

\begin{figure} 
\vspace{3.7truecm}
\includegraphics{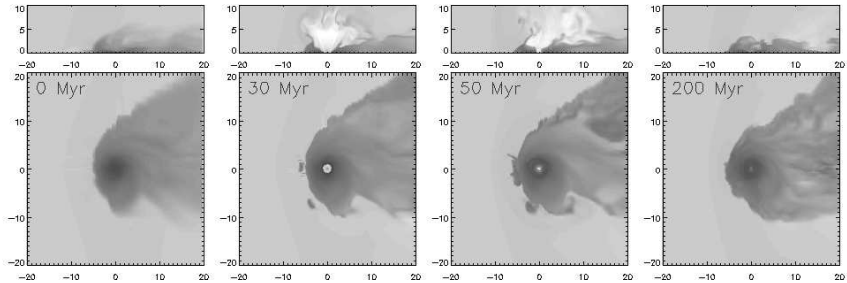}
\caption[]{Density maps on the equatorial (bottom row) and 
meridional (top row) plane, at four
different times. Axis labels are in kpc.}
\label{Fig. 1}
\end{figure}

Figure 1 shows the density maps in the equatorial and meridional plane
at four different times. The left panel refers to the time just before
the starburst onset. The ram pressure
limits the radial extent of the ISM to $r\approx 5-10$ kpc and generates the
classic gas tail. The mass in the galactic region, however, has not been 
eroded by the stripping. 

The other panels of Figure 1 shows the density at the end of the energy
injection phase (30 Myr), at $t=50$ Myr, and at a late time $t=200$ Myr.
The superbubble breaks out injecting the metal-rich gas in the ICM. 
We accurately follow the evolution of the SN ejecta as a separate fluid.
In Figure 2 are plotted the time (0-500 Myr) variations of the ISM 
mass and the 
SN ejecta fraction in the galactic region. 
We find that the starburst
is inefficient in removing the ISM. The gas mass slightly decreases
for $0<t<100$ Myr, then increases again and actually more gas is present
at the final time than at the time of the starburst onset. Shortly
after the SN activity ceased, a radial flow fills the central hole,
the initial ISM distribution is approximately recovered, and the galaxy 
might be ready for a second starburst.

The SN ejecta behaves differently. At any time only a minor fraction
of the metal-rich material is present in the galactic region. 
Some ejecta flows back in with the ISM flow when the superbubble collapse, 
but at the end of the simulation only $\sim 3$ \% of the total metals 
are found inside the galaxy.
In the model presented here the ram pressure has little effect
on the evolution of the ISM and metal-rich material 
in the galactic central region (see Figure 2).
While the ejecta is carried downstream at large distance by the moving ICM,
for a galaxy at rest most of the ejecta is dispersed in the ICM at large 
heights ($\sim 30$ kpc) over the galaxy. 
For larger ram pressures, or smaller galaxies, the ISM is significantly
modified (or completely removed) by the interaction with the ICM. 
When this happens,
galactic winds become more efficient in blowing out ISM and metals,
and a relatively weak starburst may disrupt the ISM of the parent galaxy.
A large grid of ram pressure+wind models will be described in detail
in Marcolini et al. (in preparation).

\begin{figure}
\vspace{3.6truecm}
\includegraphics{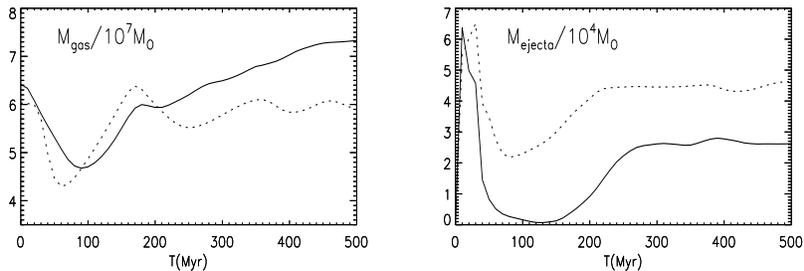}
\caption[]{Variation of the gas mass (left) and the metal-rich ejecta mass
(right) vs. time. Solid line are for the galaxy subject to ram pressure
stripping, dashed lines represent the galaxy at rest in the ICM. The total amount
of ejecta produced by the burst after 30 Myr is $9 \times 10^5 \, M_{\odot}$.  }
\label{Fig. 2}
\end{figure}


\end{article}
\end{document}